\documentclass{article}
\usepackage{amscd,amsmath,amssymb,bbm}
\usepackage{amsfonts}
\def\theequation{\arabic{section}.\arabic{equation}}

\newcommand{\p}[1]{(\ref{#1})}

\newcommand{\bD}{{\overline{\strut D}}{}}

\newcommand{\bxi}{{\bar\xi}}
\newcommand{\brho}{{\bar\rho}}
\newcommand{\bpsi}{{\bar\psi}}

\newcommand{\bPsi}{{\overline \Psi}{}}
\newcommand{\bt}{{\bar\theta}}

\newcommand{\be}{\begin{equation}}
\newcommand{\ee}{\end{equation}}
\newcommand{\bea}{\begin{eqnarray}}
\newcommand{\eea}{\end{eqnarray}}
\newcommand{\ba}{\begin{array}}
\newcommand{\ea}{\end{array}}

\newcommand{\eps}{\varepsilon}

\newcommand{\nn}{\nonumber}

\def\={\ =\ }

\begin{document}
\thispagestyle{empty}
\vspace{2cm}
\begin{flushright}
%Draft 7 \\
%16.11.2009\\
\end{flushright}\vspace{2cm}
\begin{center}
{\Large\bf Three dimensional N=4 supersymmetric mechanics with Wu-Yang monopole}
\end{center}
\vspace{1cm}

\begin{center}
{\large\bf Stefano Bellucci${}^{a}$, Sergey Krivonos${}^{b}$ and Anton Sutulin${}^{b}$ }
\end{center}

\begin{center}
${}^a$ {\it
INFN-Laboratori Nazionali di Frascati,
Via E. Fermi 40, 00044 Frascati, Italy} \vspace{0.2cm}

${}^b$ {\it
Bogoliubov  Laboratory of Theoretical Physics, JINR,
141980 Dubna, Russia} \vspace{0.2cm}

\end{center}
\vspace{2cm}

\begin{abstract}
\noindent
We propose Lagrangian and Hamiltonian formulations of a $N=4$ supersymmetric three-dimensional isospin-carrying particle
moving in the non-Abelian field of a Wu-Yang monopole and in some specific scalar potential. This additional potential is completely
fixed by $N=4$ supersymmetry and in the simplest case of flat metrics it coincides with that which provides the existence
of the Runge-Lenz vector for the bosonic subsector. The isospin degrees of freedom are described on the Lagrangian level by bosonic
auxiliary variables forming $N=4$ supermultiplet with additional, also auxiliary fermions.

Being quite general, the constructed systems include such interesting cases as $N=4$ superconformally invariant systems with Wu-Yang monopole, 
the  particles living in the
flat $\mathbb{R}{}^3$ and in the $\mathbb{R}\times \mathbb{S}{}^2$ spaces and interacting with the monopole, and also the particles moving on three-dimensional sphere and pseudo-sphere with Wu-Yang monopole sitting in the center.

The superfield Lagrangian description of these systems is so simple that one could wonder to see how all couplings and the proper
coefficients arise while passing to the component action.
\end{abstract}

\newpage
\setcounter{page}{1}

\setcounter{equation}0
\section{Introduction}
The most natural  description of quantum mechanics of a isospin-carrying particle in the background of magnetic monopole \cite{mm}
is achieved within the Hamiltonian formalism. In this setup the basic equations are Wong's equations \cite{wong} which are equations of motion
in the non-relativistic mechanics of an isospin particle in a background Yang-Mills field. Being quite enough for the bosonic particles
cases, this approach meets a lot of problems when we attempt to add supersymmetry. Indeed, the Lagrangian description in terms of
superfields is a natural framework for supersymmetric theories, especially in the case of extending $(N\geq 2)$ supersymmetries. Therefore,
to enjoy all power of the superspace approach one has to switch to the Lagrangian formalism.
The main questions in such a description are to which supermultiplet the isospin degrees of freedom belong to and how to write the corresponding
action? Clearly enough, to clarify these points one has to treat the isospin vector as a composite one. The rather old idea is to construct
isospin currents $I^a$ from the physical fermions presented in the theory as $I^a\sim (\psi\, \sigma^a \bpsi)$ (see e.g.\cite{{VH},{BSSW}} and refs. therein). The fermions $\psi$ appear  in the theory together with auxiliary bosons and after quantization $I^a$ obeys a proper algebra
(the $su(2)$ one in the present case). The Lagrangian for fermions has a standard form and coupling with physical bosons and fermions
can be also done in a standard manner. So, everything is nice besides additional unpleasant features arising from the fermionic nature of the building blocks - $\psi$ variables. For example, in the case of $su(2)$ symmetry any quadratic combination $I^a I^b$ will be proportional to $\delta^{ab}$,
while the triple product $I^a I^b I^c$ will be identically equal to zero. The evident way out of such problem is to construct the isospin currents from
bosonic variables $I^a\sim (u\, \sigma^a \bar u)$. Then, if we insist that $I^a$ have to obey $su(2)$ algebra after quantization, these bosonic
variables $u,{\bar u}$ have to enter the Lagrangian with first-order in time derivatives kinetic term as fermions $\psi$ did. Again, there are
no problems with the realization of this idea in the purely bosonic case (see e.g. \cite{MP2}). Recently, such description has been successfully used
to obtain a Lagrangian formulation of a isospin particle interacting with the Yang monopole within the second Hopf map \cite{brazil}. Nevertheless,
using the bosonic variables as the building blocks for isospin currents yields back the question about accompanying fermions and the action.

A first solution of these problems has been proposed in \cite{FIL}, where auxiliary bosonic and fermionic degrees of freedom constitute an
auxiliary gauge supermultiplet. Then this idea has been used  for the construction of $N=4$ superconformal mechanics with isospin degrees of freedom
\cite{FIL1}. This approach is rather involved and a more economical setup, in which one may reproduce (at least) the one particle case, has been developed in \cite{KL}. Both these approaches suffer from the same problem -- it is not immediately clear how to extend them to the supersymmetric systems with more than one physical bosonic component. Such an extension is absolutely necessary to describe isospin-carrying particles.

A third approach, which we will advocate here \cite{BK1}, used the ordinary superspace and ordinary superfields. Its key feature, which makes
fermions into auxiliary components and gives first order in time derivatives kinetic term for bosonic variables, is just a specific coupling
between two $N=4$ supermultiplets involved in the game. This coupling reads extremely simple in the superspace as
$$
S_c=\int dt d^4\theta  X\; \Psi^\alpha \bPsi_\alpha ,
$$
where $X$ and $\Psi$ are a set of properly constrained $N=4$ superfields (see next Sections for the details). Despite its simplicity, the
action $S_c$, being supplied by the additional action for the physical supermultiplet $X$, did everything. One may enjoy to see how
all needed terms will appear one by one while passing to the component action. But if we are treating the scalar bosonic superfield $X$ as an independent one, as it was done in the paper \cite{BK1}, there is no hope to describe the isospin particles. In the present Letter we proposes to consider $X$ as a composite superfield, constructed from $N=4$ tensor supermultiplet $V^{ij}$ in a rather simple way as
$$
X=\frac{1}{\sqrt{V^{ij}V_{ij}}}.
$$
Just this composite nature of the basic superfield $X$ provides many nice features of our model. It describes the three-dimensional
particle with isospin moving in the field of a Wu-Yang monopole \cite{WY} equipped with a specific potential term.
This additional potential is completely
fixed by $N=4$ supersymmetry and in the simplest cases of flat metrics it coincides with that which provides the existence
of the Runge-Lenz vector for the bosonic subsector \cite{{H1},{HN}}. The isospin degrees of freedom are described  by bosonic
auxiliary variables forming a $N=4$ supermultiplet with additional, also auxiliary fermions. All these features appeared automatically
due to a specific coupling in the extremely simple action $S_c$. In the next two Sections we present the detailed description of our
system in terms of superfields. The reader who is not interested in the superfield approach may immediately pass to the components
Lagrangian \p{finaction} which is our main result. In Section 4 we develop the Hamiltonian approach and present $N=4$ supercharges
in an explicit form. Section 5 collects many particulary interesting (at least for us) cases, including $N=4$ superconformally invariant systems
with Wu-Yang monopole, particles living in the
flat $\mathbb{R}{}^3$ and in the $\mathbb{R}\times \mathbb{S}{}^2$ spaces and interacting with the monopole, and also particles moving on three-dimensional sphere and pseudo-sphere with Wu-Yang monopole sitting in the center.
Finally, we discuss some unsolved problems and possible extensions of the present construction.

\setcounter{equation}0
\section{Preliminaries}
One of the possibilities to invent spin-like variables in the Lagrangian of some supersymmetric mechanics is to
introduce a specific coupling of the basic supermultiplet with another ``auxiliary'' superfield which contains
these spin-variables together with auxiliary fermions. The most essential feature of such a coupling is a first-order
kinetic term for the spin-variables. The first realization of this idea has been proposed in \cite{FIL1} where the spin-variables
and auxiliary fermions seat in the auxiliary gauge supermultiplet. Another approach, which we will follow in this Letter,
has been elaborated in \cite{BK1}. There, spin-variables and auxiliary fermions were put in a doublet of fermionic superfields
$\Psi{}^\alpha,\bPsi_\alpha$ subjected to the irreducible conditions \cite{IKL1} \footnote{If we
combine the spinor derivatives $D^i,\bD^i$ in the quartet of  derivatives $\nabla^{i\alpha}=\left\{ D^i, \bD^i\right\}$ then
the constraints \p{Psi} acquire the familiar form
$\nabla^{i(\alpha}\Psi^{\beta)}=0$}
\be\label{Psi} D^i \Psi{}^1=0, \; D^i
\Psi{}^2+\bD^i \Psi{}^1=0, \; \bD_i \Psi{}^2=0.
\ee
Here, we introduced the N=4 spinor covariant derivatives as
\be
D^i=\frac{\partial}{\partial \theta_i}+i\bt{}^i\partial_t,\;
\bD_i=\frac{\partial}{\partial \bt{}^i}+i\theta_i\partial_t,\qquad \left\{ D^i,\bD_j\right\}=2i \delta^i_j \partial_t.
\ee
The constraints \p{Psi} leave in the superfields $\Psi{}^\alpha,\bPsi_\alpha$ four fermionic and four auxiliary bosonic components
\be\label{compPsi}
\psi^\alpha= \Psi{}^\alpha|,\;\bpsi_\alpha= \bPsi{}_\alpha|,\qquad u^i= -D{}^i\bPsi{}^2|, \;
{\bar u}_i=  \bD_i \Psi{}^1|,
\ee
where, as usual, $|$ in the r.h.s. denotes the $\theta=\bt=0$ limit.

Following \cite{BK1} let us introduce the coupling of $\Psi$ supermultiplet with some arbitrary N=4 superfield $X$ subjected to the constraints
\be\label{X}
D^i D_i X =  0,\quad \bD_i \bD{}^i X = 0,\quad \bigl[ D^i, \bD_i \bigr] X=0,
\ee
as
\be\label{actionminus1}
S_c=-\frac{1}{32}\int dt d^4\theta  X\; \Psi^\alpha \bPsi_\alpha .
\ee
Being rewritten in terms of the components of $\Psi$ \p{compPsi} and $X$ defined as\footnote{We defined symmetrization over indices as
$a_{(ij)}\equiv \frac{1}{2}\left( a_{ij}+a_{ji}\right)$. The raising and lowering of the indices goes as $A^i=\epsilon^{ij}A_j, A_i=\epsilon_{ij}A^j$, and thus $\epsilon_{ij}\epsilon^{jk}=\delta_i^k$.}
\be\label{compX}
 x= X|,\; A_{ij} = A_{(ij)}= \frac{1}{2}\left[ D_i,\bD_j\right] X|,\qquad \eta^i= -iD^i X|,\;
\bar\eta{}_i= -i\bD_i X|,
\ee
the action \p{actionminus1} reads
\bea\label{action0}
S_c&=&
\int dt\left[ -x \left({\dot \psi}{}^1{\dot\bpsi}{}^2-{\dot \psi}{}^2{\dot\bpsi}{}^1\right)-\frac{i}{4} x \left( {\dot u}{}^i {\bar u}_i-
u^i\dot{\bar u}_i\right)+\frac{1}{4}A_{ij}u^i{\bar u}{}^j+\right.\nn \\
&&\left. \frac{1}{2}\eta_i\left({\bar u}{}^i \dot\bpsi{}^2+u^i \dot\psi{}^2\right)+\frac{1}{2}\bar\eta{}^i\left(u_i \dot\psi{}^1+{\bar u}_i\dot\bpsi{}^1\right)
\right].
\eea
The crucial observation of \cite{BK1} is that in this action
the fermionic fields $\psi^\alpha$ and $\bpsi{}_\alpha$ appeared only
through the time derivatives and therefore one may replace
these time derivatives  by new fermionic fields $\rho^\alpha$ and $\brho{}_\alpha$ as
\be\label{xi}
\rho^\alpha= {\dot\psi}{}^\alpha, \qquad \brho_\alpha=\dot\bpsi{}_\alpha.
\ee
This is nothing but the reduction from the $(0,4,4)$
supermultiplet to the auxiliary $(0,0,4,4)$ one
\cite{GR,root,SM}. This reduction is compatible with $N=4$
supersymmetry and the  transformation properties of the new
fermions $\rho^\alpha,\brho{}_\alpha$ forming new $N=4$ supermultiplet together with the bosonic components $u^i, {\bar u}_i$, read
\be\label{ad2}
\delta \rho^1=-\bar\epsilon{}^i \dot{\bar u}_i,\;
\delta\rho^2=\epsilon_i\dot{\bar u}{}^i,\quad \delta
u^i=-2i\epsilon^i\brho{}^1+2i\bar\epsilon{}^i\brho{}^2,\; \delta
{\bar u}_i=-2i\epsilon_i\rho{}^1+2i\bar\epsilon_i\rho{}^2.
\ee
Now one may easily check that the action
\bea\label{action1}
S_c&=
&\int dt\left[ -x\left({ \rho}{}^1{\brho}{}^2-{ \rho}{}^2{\brho}{}^1\right)-\frac{i}{4} x \left( {\dot u}{}^i {\bar u}_i-
u^i\dot{\bar u}_i\right)+\frac{1}{4}A_{ij}u^i{\bar u}{}^j+\right.\nn\\
&& \left. \frac{1}{2}\eta_i\left({\bar u}{}^i \brho{}^2+u^i
\rho{}^2\right)+\frac{1}{2}\bar\eta{}^i\left(u_i \rho{}^1+{\bar
u}_i\brho{}^1\right) \right],
\eea
is invariant under \p{ad2}, provided the components of the $X$ supermultiplet transform in a
standard way as
\be\label{ad4} \delta
x=-i\epsilon_i\eta^i-i\bar\epsilon{}^i\bar\eta_i,\quad
\delta\eta{}^i=-\bar\epsilon{}^i{\dot x}-i\bar\epsilon{}^j
A^i_j,\; \delta \bar\eta_i=-\epsilon_i{\dot x}+i\epsilon_j
A_i^j,\quad \delta A_{ij} = -\epsilon_{(i}\dot\eta_{j)} +
\bar\epsilon_{(i}\dot{\bar\eta}{}_{j)}.
\ee
In the action \p{action1}
the fermionic fields $\rho^\alpha,\brho{}_\alpha$  are auxiliary ones and thus they can be
eliminated  by their equations of motion
\be\label{ad5a}
\rho^1=\frac{1}{2x}\eta_i{\bar u}{}^i, \qquad
\rho^2=-\frac{1}{2x}\bar\eta{}^i{\bar u}_i.
\ee
{}Finally, the action describing the interaction of $\Psi$ and $X$ supermultiplets acquires the form:
\be\label{action2}
S_c=\frac{1}{4}\int dt\left[
 -i x \left( {\dot u}^i {\bar u}{}_i- u^i\dot{\bar u}{}_i\right)+
 A_{ij}u^i{\bar u}{}^j+
\frac{1}{x} \eta_i\bar\eta_j\left( u^i {\bar u}{}^j+ u^j{\bar u}{}^i\right)
\right].
\ee
Thus, we see that from fermionic superfields $\Psi{}^a$ survive only bosonic
components $u^i,{\bar u}_i$ entering the action with kinetic term linear in time-derivatives. After quantization
these variables become purely internal degrees of freedom.

The next step is to extend the set of bosonic variables which have to interact with "spin"-variables by specializing
the supermultiplet $X$ and its self-coupling.

\setcounter{equation}0
\section{Coupling with tensor supermultiplet}
To be meaningful the action \p{action2} describing coupling of $\Psi$ and $X$ supermultiplets, has to be extended by the
action for the supermultiplet $X$ itself. Clearly, the most general action of such type reads
\be\label{action3}
S_x = -\frac{1}{32}\int dt d^4\theta
{\cal F}(X),
\ee
where ${\cal F}(X)$ is an arbitrary function of $X$. In terms of components \p{compX} the action \p{action3} has the form
\be\label{action4}
S_x=\frac{1}{8}\int dt\left[F'' {\dot x}{}^2-\frac{1}{2} F''
A^{ij}A_{ij}+i F''\left(\dot\eta{}^i\bar\eta_i-
\eta^i\dot{\bar\eta}_i\right)+F'''\eta^i\bar\eta{}^jA_{ij}-\frac{1}{4}F^{(4)}\eta^i\eta_i\bar\eta_j\bar\eta{}^j\right],
\ee
where
\be
F(x) = {\cal F}(X)|.
\ee
If we consider the superfield $X$ as an independent superfield then the sum of the actions
\p{action3} and  \p{actionminus1}
\be\label{GA}
S=S_x + S_c
\ee
describes $N=4$ supersymmetric mechanics with one physical boson $x$ and four physical fermions
$\eta^i, {\bar\eta}_j$ interacting with spin-variables $u^i, {\bar u}_i$. Just this system has been considered in \cite{{BK1}, {FIL}, {FIL1}}.
The main idea of the present Letter is to consider the superfield $X$ as a composite one, constructed from another ``basic'' superfields containing
more physical bosons. Let us choose as the ``basic'' superfields the triplet of bosonic $N=4$ superfields $V^{ij}=V^{ij}$ subjected to
the constraints
\be\label{V}
D^{(i}V^{jk)}=\bD{}^{(i}V^{jk)}=0, \qquad \left( V^{ij}\right)^\dagger = V_{ij}.
\ee
The constraints \p{V} leave in $V^{ij}$ the following independent components:
\be\label{v}
V^{ij}|=v^{ij},\quad
D^i V^{kn}| = -\frac{1}{3} (\eps^{ik} \lambda^n + \eps^{in} \lambda^k), \;
\bD_i V^{kn}| = \frac{1}{3} (\delta^k_i \bar \lambda^n + \delta^n_i \bar \lambda^k),\quad
D^i \bar D^j V_{ij}|=A.
\ee
Thus its off-shell component field content is $(3,4,1)$, i.e. three physical $v^{ij}$ and one auxiliary $A$
bosons and four fermions $\lambda^i, \bar\lambda_i$ \cite{{v1},{v1a}}.

Now one may check that the composite superfield
\be\label{XV}
X=\frac{1}{\sqrt{V^2}} \equiv \frac{1}{\sqrt{V^{ij}V_{ij}}}
\ee
obeys \p{X} in virtue of \p{V}. The main novel feature of the representation \p{XV} is that all components of the $X$ superfield,
i.e. the physical boson $x$, fermions $\eta^i, \bar\eta_i$ and auxiliary fields $A^{ij}$ \p{compX}
are expressed through the components of $V^{ij}$ supermultiplet \p{v} as
\bea\label{rel1}
&& \eta^i =  - \frac{2i}{3} x^3 (v^i_n \lambda^n), \quad
\bar\eta_i =  \frac{2i}{3} x^3 (v_{in} \bar \lambda^n),\qquad x=\frac{1}{\sqrt{v^2}}, \nn \\
&&
A_{ij} =-i x^3 (v_{in} \dot v_j^n + v_{jn} \dot v_i^n) - \frac{1}{3} x^3 v_{ij} A +
\frac{2}{9} x^3 (\lambda_i \bar \lambda_j + \lambda_j \bar \lambda_i)
- \frac{4}{3} x^5 v_{ij} v_{kn} \lambda^k \bar \lambda^n.
\eea
In what follows we prefer to work with $\eta^i, \bar\eta_i$ fermions and therefore the suitable expression for $A_{ij}$ will be
\be\label{rel2}
A_{ij} =-i x^3 (v_{in} \dot v_j^n + v_{jn} \dot v_i^n) - \frac{1}{3} x^3 v_{ij} A
+ 2 x ( v_{ik} v_{jn} + v_{jk} v_{in} - 3 v_{ij} v_{kn})\eta^k \bar \eta^n.
\ee

Now we substitute the fields $A_{ij}$ in the  action  \p{GA}, eliminate the auxiliary field $A$ by its
equation of motion and obtain the following action:
\bea\label{lag1}
S &=& \int dt \left[ \frac{x^4}{8} F'' \dot v^{kn} \dot v_{kn} +
\frac{i}{8} F'' (\dot \eta^i \bar \eta_i - \eta^i \dot{\bar \eta}_i)
- \frac{ix^2}{4} F'' \bigg [ 1 + \frac{x}{2} \frac{F'''}{F''}\bigg ] (v_{in} \dot v^n_j + v_{jn} \dot v^n_i) \eta^i \bar \eta^j \right.\nn \\
&-& \frac{i}{4} x (\dot u^i \bar u_i - u^i \dot{\bar u}_i)
- \frac{ix^3}{4} (v_{in} \dot v^n_j + v_{jn} \dot v^n_i) u^i \bar u^j \nn\\
&+& \frac{x^2}{4 F''} v_{ij} v_{kn} u^i \bar u^j u^k \bar u^n
+\frac{1}{4x} (u_i \bar u_j + u_j \bar u_i)
 \eta^i \bar \eta^j \nn\\
&+& \frac{x}{2} (v_{ik} v_{jn} + v_{jk} v_{in} - v_{ij} v_{kn}) \eta^k \bar \eta^n u^i \bar u^j
+ \frac{x^2}{4} \frac{F'''}{F''} v_{ij} v_{kn} \eta^k \bar \eta^n u^i \bar u^j \nn\\
&-& \left.\frac{1}{8x^2} \bigg [ F'' + x F''' - \frac{x^2}{8} \frac{(F''')^2}{F''} + \frac{x^2}{4} F^{(4)} \bigg ] \eta^2 \bar \eta^2\right].
\eea
As the last step, to simplify the action \p{lag1}, we pass to the new variables defined as
\bea\label{fincomp}
&&
v^a = -\frac{i}{\sqrt{2}}\, (\sigma^a)^j_i v^i_j \qquad \Longrightarrow \qquad
x = \frac{1}{(v^a v^a)^{1/2}} \nn\\
&&
\omega^i = u^i\, \sqrt{x}\,, \quad
\bar \omega_i = \bar u_i\, \sqrt{x}\,, \qquad
\xi^i = \eta^i\, \frac{\sqrt{F''}}{2}\,, \quad
\bar \xi_i = \bar \eta_i\, \frac{\sqrt{F''}}{2}\,,
\eea
where $\sigma^a$-matrices are chosen to satisfy $\left[ \sigma^a, \sigma^b\right]=2i \epsilon^{abc}\sigma^c$. In these variables the
action \p{lag1} reads
$$S =\int dt\left[ \frac{x^4}{8} G \dot v^a \dot v_a +
\frac{i}{2} (\dot \xi^i \bar \xi_i - \xi^i \dot{\bar \xi}_i)
- \frac{i}{4} (\dot \omega{}^i \bar \omega_i - \omega^i \dot{\bar \omega}_i)+i\frac{x^2}{2}\eps^{abc} v^a \dot v^b\left( I^c-2
\left( 1 + \frac{x}{2} \frac{G'}{G} \right)\Sigma^c\right)
 \right.$$
\be\label{finaction}
+ \left. \frac{1}{2 G} (v^a I^a)^2-
\frac{2}{ G} \left( 1 + \frac{x}{2} \frac{G'}{G} \right) (v^a I^a) (v^b \Sigma^b)
- \frac{4}{3 x^2 G^2} \left( G + x G' - \frac{x^2(G')^2}{8G} + \frac{x^2}{4} G''
\right) \Sigma^a \Sigma^a\right],
\ee
where
\be\label{findef}
G(x)\equiv F''(x), \qquad I^a = \frac{i}{2}(\sigma^a)^j_i \omega^i \bar \omega_j, \quad \Sigma^a = -i(\sigma^a)^j_i \xi^i \bar \xi_j\,.
\ee
One may check that the action \p{finaction} is perfectly invariant under $N=4$ supersymmetry. We present explicit $N=4$ supersymmetry transformations of the physical components in Appendix \p{susycomp}.

The action \p{finaction} is the main result of this Letter. It describes the $N=4$ supersymmetric three-dimensional isospin particles moving in the magnetic field of some monopole. It is not too hard to understand that the corresponding monopole will be of Wu-Yang type \cite{WY}. Indeed, from
\p{finaction} we see that the bosonic part of the vector potential of the monopole reads
\be\label{WY}
{\cal A}{}^a = -\frac{i}{2 v^2}\epsilon^{abc}v^b I^c.
\ee
This is just the potential of the Wu-Yang monopole if we will be able to treat $I^a$, defined in \p{findef}, as the proper isospin matrices\footnote{To be a solution of $su(2)$ Yang-Mills equations the isospin matrices $I^a$ entering the potential ${\cal A}{}^a$ have to commute exactly as in \p{S}}.
This could be done in the Hamiltonian formalism which we will consider in the next Section. Another interesting feature of the action \p{finaction}
is appearance of the potential term
\be\label{po1}
V= \frac{1}{2 G} (v^a I^a)^2.
\ee
In the case of the flat metrics case, corresponding to  $G=4/x^4$, it reduces to
\be\label{po1a}
V_{WY}= \frac{1}{8 (v^2)^2} (v^a I^a)^2.
\ee
In this form it completely coincides with the potential found in \cite{{H1},{HN}} which is absolutely needed to provide the isospin-particle moving
in the field of a Wu-Yang monopole with the conserved Runge-Lenz vector. In our approach this potential appears automatically, accompanying the
Wu-Yang monopole potential in the $N=4$ supersymmetric action \p{finaction}.

\setcounter{equation}0
\section{Hamiltonian and Supercharges}
In order to find the classical Hamiltonian, we follow the standard procedure for quantizing a system
with bosonic and fermionic degrees of freedom. From the action \p{finaction} we define the momenta
$P^a, \pi_i, \bar\pi{}^i, p_i,{\bar p}{}^i$ conjugated to $v^a, \xi^i, \bxi_i, \omega^i$ and
$\bar\omega_i$, respectively, as
\bea\label{momenta}
&&
P^a = \frac{x^4}{4} G \dot v^a - \frac{i}{2} x^2 \eps^{abc} v^b \left( I^c -2 \left( 1 + \frac{x}{2} \frac{G'}{G} \right)  \Sigma^c\right)\,, \nn \\
&&
\pi_i = \frac{i}{2}\, \bar \xi_i, \quad
\bar \pi^i = \frac{i}{2}\, \xi^i, \quad
p_i = -\frac{i}{4}\, \bar \omega_i, \quad
\bar p^i = \frac{i}{4}\, \omega^i,
\eea
and introduce Dirac brackets for the canonical variables
\be\label{PB}
\{v^a,P^b \} = \delta^{ab}, \qquad \{ \xi^i, \bar \xi_j \} = i \delta^i_j, \quad
\{ \omega^i, \bar \omega_j \} = 2i \delta^i_j.
\ee
Now one may check that the supercharges
\bea\label{SCh}
Q^i &=& \frac{i}{x \sqrt{G}}\, v^a \bigg [ \delta^i_j P^a + i \eps^{abc} P^b (\sigma^c)^i_j \bigg] \bar \xi^j
+ \frac{i}{2 x \sqrt{G}}\,  (\sigma^a)^i_j \bar \xi^j I^a
- \frac{i}{6}\frac{G'}{G\sqrt{G}}  (\sigma^a)^i_j \bar \xi^j \Sigma^a\,, \nn\\
\bar Q_i &=& \frac{i}{x \sqrt{G}}\, v^a \bigg [ \delta^j_i P^a - i \eps^{abc} P^b (\sigma^c)^j_i \bigg] \xi_j
- \frac{i}{2 x \sqrt{G}}\,  (\sigma^a)^j_i \xi_j I^a
+ \frac{i}{6} \frac{G'}{G\sqrt{G}}  (\sigma^a)^j_i \xi_j \Sigma^a\,,
\eea
and the Hamiltonian
\bea\label{HAM}
H &=& \frac{2}{x^4 G} \widehat{P}{}^a \widehat{P}{}^a
- \frac{1}{2G} (v^a I^a)^2
+ \frac{2}{G} \left( 1 + \frac{x}{2} \frac{G'}{G}\right)(v^a I^a) (v^b \Sigma^b) \nn\\
&+& \frac{4}{3 x^2\ G^2} \left( G + x G' - \frac{x^2}{8} \frac{(G')^2}{G} + \frac{x^2}{4} G''
\right) \Sigma^a \Sigma^a\,,
\eea
where
\be
\widehat{P}{}^a= P^a - i x^2 \left( 1 + \frac{x}{2} \frac{G'}{G} \right) \eps^{abc} v^b \Sigma^c
+\frac{i}{2} x^2\eps^{abc}  v^b I^c
\ee
form the $N=4$ Poincar\`{e} superalgebra
\be\label{poincare}
\left\{ Q^i,  {\bar Q}_j\right\} = \frac{i}{2}\, \delta^i_j\, H .
\ee
{}From the explicit form of the Hamiltonian one may see that spin-variables enter it only through the three
dimensional vector $I^a$. This vector $I^a$ commutes with everything, excluding itself, with
which it forms a $su(2)$ algebra with respect to the brackets \p{PB}
\be\label{S}
\left\{ I^a, I^b\right\} = 2i \epsilon^{abc} I^c.
\ee
In addition, the module $I^2=I^a I^a$ commutes with the  Hamiltonian. Thus, the relations \p{S}
define the classical isospin matrices with fixed isospin modulus. At the same time the fermions enter the
Hamiltonian only through the combination $\Sigma^a$ which also obeys the brackets
\be\label{Sigma}
\left\{ \Sigma^a, \Sigma^b\right\} = 2i \epsilon^{abc} \Sigma^c,
\ee
thus providing a description for the fermionic spin degrees of freedom.

Therefore, we conclude that the Hamiltonian \p{HAM} indeed describes the motion of the $N=4$ supersymmetric
isospin particle in the field of a Wu-Yang monopole and a specific potential.

\setcounter{equation}0
\section{Cases of interest}
Up to now our consideration was  quite general, and the superpotential ${\cal F} (X)$ \p{action3} as well as induced metrics in the configuration
space $x^4 G(x)/8$ were arbitrary. Let us now specify the superpotential to get the cases of special interest.
\subsection{Superconformal invariant models}
From the general considerations presented in \cite{{BK1},{IKL2}} it follows that our models, in which we specified the  superfield $X$ to be
 a composite one, possess invariance with respect to the most general
$N=4$ superconformal group in one dimension, i.e. the $D(2,1;\alpha)$  one \cite{Sorba}, provided we choose the superpotential ${\cal F}(X)$ as
\bea\label{SConf}
&& {\cal F}(X) = X^{-\frac{1}{\alpha}} \quad \Rightarrow \quad G(x)=\frac{1+\alpha}{\alpha^2}x^{-2-\frac{1}{\alpha}}, \qquad \alpha \neq 0, -1 \nn \\
&& {\cal F}(X) = X \log X \quad \Rightarrow \quad G(x)= \frac{1}{x} ,\qquad \alpha=-1.
\eea
The corresponding three-dimensional geometry is a 3-dimensional cone $C(B)$ over the base manifold
$B=\mathbb{S}^2$ of radius $\frac{1}{4\alpha^2}$ \cite{IKL1}. At $\alpha=\pm 1/2$ one recovers flat space $\mathbb{R}^3$, while at any other $\alpha$ we will have a curved manifold. Such a conical geometry is
typical for the bosonic sectors of superconformal theories in diverse dimensions \cite{{GiR}, {BNY1}}.

Among the superconformally invariant theories the flat metrics case with $\alpha=1/2$ is rather interesting due to very simple form
of the Lagrangian (we choose $G(x)=\frac{4}{x^4}$):
\bea\label{half}
S_{\alpha = \frac{1}{2}} &=&\int dt\left[ \frac{1}{2} \dot v^a \dot v^a +
\frac{i}{2} (\dot \xi^i \bar \xi_i - \xi^i \dot{\bar \xi}_i)
+ \frac{i}{4} (\dot \omega_i \bar \omega^i - \omega_i \dot{\bar \omega}^i)+i\frac{1}{2\, v^2}\eps^{abc} v^a \dot v^b\left( I^c+2
\Sigma^c\right) \right .\nn \\
&&
+ \left. \frac{1}{8\,(v^2)^2} (v^a I^a)^2+
\frac{1}{2\,(v^2)^2}  (v^a I^a) (v^b \Sigma^b)\right], \qquad v^2 \equiv v^a v^a.
\eea
As we can see, the four-fermionic term disappeared from the action in this case.

The second choice for the flat metrics in the configuration space $(\alpha=-1/2)$ is achieved with $G=const=4$. In this case the kinetic
term for $v^a$ components reads
\be\label{flat1}
L_{kin}= \frac{x^4}{2} {\dot v}{}^a {\dot v}{}^a = \frac{x^4}{2}\left( \frac{{\dot x}{}^2}{x^4}+\frac{1}{x^2}{\dot{\hat v}}{}^a {\dot{\hat v}}{}^a\right)\equiv \frac{1}{2} \dot{\tilde v}^a \dot{\tilde v}^a , \qquad {\hat v}{}^a{\hat v}{}^a=1,\quad {\tilde v}{}^a{\tilde v}{}^a=x^2.
\ee
With these bosonic variables the action \p{finaction} reads
$$S_{\alpha = -\frac{1}{2}} =\int dt\left[ \frac{1}{2} \dot{\tilde v}^a \dot{\tilde v}^a +
\frac{i}{2} (\dot \xi^i \bar \xi_i - \xi^i \dot{\bar \xi}_i)
+ \frac{i}{4} (\dot \omega_i \bar \omega^i - \omega_i \dot{\bar \omega}^i)+\frac{i}{2\, {\tilde v}{}^2}\eps^{abc} {\tilde v}^a \dot{\tilde v}{}^b\left( I^c-2\Sigma^c\right)
 \right.$$
\be\label{minushalf}
+ \left. \frac{1}{8\, ({\tilde v}{}^2)^2} ({\tilde v}{}^a I^a)^2-
\frac{1}{2\, ({\tilde v}{}^2)^2}  ({\tilde v}{}^a I^a) ({\tilde v}{}^b \Sigma^b)
- \frac{1}{3 \, {\tilde v}{}^2}  \Sigma^a \Sigma^a\right].
\ee
In this case the four-fermionic term is needed for $N=4$ supersymmetry.

The actions for the remaining values of $\alpha$ do not have peculiarities or/and simplifications.

\subsection{$\mathbb{R}\times \mathbb{S}^2$ case}
Another selected case corresponds to the choice
\be
{\cal F}(X)=-4\log X\quad \Rightarrow G=\frac{4}{x^2}.
\ee
With this choice the kinetic term for the bosonic $v^a$ variables in the action \p{finaction} acquires the form
\be
\frac{x^2}{2}\,  \dot v^a\, \dot v_a = \frac{1}{2} \left(  \frac{{\dot x}^2}{x^2} + {\dot{\hat v}}{}^a\, {\dot{\hat v}}{}^a \right), \quad
{\hat v}{}^a {\hat v}{}^a=1,
\ee
and thus we meet the $\mathbb{R}\times \mathbb{S}^2$ geometry in the bosonic sector. The full action now reads
\be\label{RS2}
S_{\mathbb{R}\times \mathbb{S}^2} =\int dt\left[ \frac{1}{2} \left( \frac{{\dot x}^2}{x^2} + {\dot{\hat v}}{}^a\, {\dot{\hat v}}{}^a \right) +
\frac{i}{2} (\dot \xi^i \bar \xi_i - \xi^i \dot{\bar \xi}_i)
+ \frac{i}{4} (\dot \omega_i \bar \omega^i - \omega_i \dot{\bar \omega}^i)+\frac{i}{2}\eps^{abc} {\hat v}{}^a \dot{\hat v}{}^b I^c+
\frac{1}{8} ({\hat v}{}^a I^a)^2\right].
\ee
In this action only two components of $v^a$, i.e. ${\hat v}{}^a$, interact with isospin degrees of freedom, while the fermions together
with the third physical field $x$ are free. Nevertheless, the action \p{RS2} is still invariant with respect to $N=4$ supersymmetry.
\subsection{Sphere and pseudo-Sphere}
Surely, sphere and pseudo-sphere are the best known 3-dimensional manifolds. Therefore, we will present the corresponding actions for
$N=4$ supersymmetric systems with such bosonic manifolds explicitly.
\subsubsection{Sphere $\mathbb{S}^3$}
To describe sphere $\mathbb{S}^3$ one has to choose
\be\label{s31}
{\cal F}(X) = 2 \arctan X \quad \Rightarrow \quad \frac{x^4}{8} G = \frac{1}{2\, (1+v^2)^2}.
\ee
With this choice our action \p{finaction} reads
$$S =\int dt\left[ \frac{ \dot v^a \dot v^a}{2\,(1+v^2)^2} +
\frac{i}{2} (\dot \xi^i \bar \xi_i - \xi^i \dot{\bar \xi}_i)
+ \frac{i}{4} (\dot \omega_i \bar \omega^i - \omega_i \dot{\bar \omega}^i)+\frac{1}{2 v^2}\eps^{abc} v^a \dot v^b\left( I^c+2
\frac{1-v^2}{1+v^2}\Sigma^c\right)
 \right.$$
\be\label{S3action}
+ \left. \frac{(1+v^2)^2}{8 (v^2)^2} (v^a I^a)^2+
\frac{1-(v^2)^2}{2 (v^2)^2}  (v^a I^a) (v^b \Sigma^b)
+ \left( 1-\frac{v^2}{3}\right) \Sigma^a \Sigma^a\right].
\ee
\subsubsection{Pseudo-Sphere}
To get a pseudo-sphere in the bosonic sector one has choose the superfield prepotential $\cal F$ as
\be\label{ps31}
{\cal F}(X) = X\log\left( \frac{1+X}{1-X}\right)  \quad \Rightarrow \quad \frac{x^4}{8} G = \frac{1}{2\, (1-v^2)^2}.
\ee
Now, our action \p{finaction} is simplified to be
$$S =\int dt\left[ \frac{ \dot v^a \dot v^a}{2\,(1-v^2)^2} +
\frac{i}{2} (\dot \xi^i \bar \xi_i - \xi^i \dot{\bar \xi}_i)
+ \frac{i}{4} (\dot \omega_i \bar \omega^i - \omega_i \dot{\bar \omega}^i)+\frac{1}{2 v^2}\eps^{abc} v^a \dot v^b\left( I^c+2
\frac{1+v^2}{1-v^2}\Sigma^c\right)
 \right.$$
\be\label{pS3action}
+ \left. \frac{(1-v^2)^2}{8 (v^2)^2} (v^a I^a)^2+
\frac{1-(v^2)^2}{2 (v^2)^2}  (v^a I^a) (v^b \Sigma^b)
- \left( 1+\frac{v^2}{3}\right)  \Sigma^a \Sigma^a\right].
\ee

\setcounter{equation}0
\section{Discussions and perspectives}
In this Letter we have proposed the Lagrangian and the Hamiltonian formulations of $N=4$ supersymmetric three-dimensional isospin-carrying particle
moving in the non-Abelian field of a Wu-Yang monopole and of some specific scalar potential. This additional potential is completely
fixed by $N=4$ supersymmetry and in the simplest cases of flat metrics it coincides with that which provides the existence
of the Runge-Lenz vector for the bosonic subsector. The isospin degrees of freedom are described on the Lagrangian level by bosonic
auxiliary variables forming $N=4$ supermultiplet with additional, also auxiliary fermions.

The superfield Lagrangian description of these systems is so simple that one could wonder to see how all couplings and the proper
coefficients arise one by one in the component action.

All our consideration here was a classical one. So, one of the immediate tasks is to provide a quantum description for, at least, one of
the listed systems, in order to clarify the role of $N=4$ supersymmetry. Another very interesting task, which  we completely ignored in the
present Letter, is to check the existence of the conserved Runge-Lenz vectors in our $N=4$ supersymmetric systems. The strict form
of the arising potential which is accompanying the Wu-Yang monopole and which is compatible with the existence of Runge-Lenz vectors in the bosonic sector,
looks quite promising. The related question of the existence of nonlinear supersymmetry in our systems (see e.g. \cite{MP1}) is also open.

Being quite general, the constructed systems include such interesting cases as $N=4$ superconformally invariant systems with
$D(2,1;\alpha)$ superconformal group, particles in the
flat $\mathbb{R}{}^3$ and $\mathbb{R}\times \mathbb{S}{}^2$ spaces and also particles moving on three-dimensional sphere and pseudo-sphere.
Nevertheless, there are some possible extensions of the constructed system in the following directions
\begin{itemize}
\item First of all one could consider  more general actions for a self-coupled $N=4$ tensor $V^{ij}$ supermultiplet, instead of $S_x$ \p{action3}, by treating the
 prepotential ${\cal F}$ as an arbitrary function depending on superfields $V^{ij}$. The corresponding system will include
a more complicated metric in the configuration space and more complicated couplings, as it happened in the case without isospin variables
\cite{{sm1},{IL}}.
\item Another possible extension includes additional self-coupling action for the $\Psi$ supermultiplet like
$$S_\beta \sim \beta \int dt d^4\theta  \Psi^\alpha \bPsi_\alpha,$$ with some arbitrary constant parameter $\beta$. This additional action will preserve all nice features of our system, but the potential of the monopole will change to be
    $${\cal A}{}^a_\beta = -\frac{i}{2 \sqrt{v^2}(\sqrt{v^2}+\beta)}\epsilon^{abc}v^b I{}^c$$
    instead of the Wu-Yang monopole \p{WY}.
\item While working with the tensor $V^{ij}$ supermultiplet one may introduce more complicated couplings with auxiliary $\Psi$ supermultiplet like
$$S_{ad}\sim \int dt d^4\theta \left(\beta_1 V_{ij}+ \beta_2 \frac{V_{ij}}{(V^2)^{\frac{3}{2}}}\right) \Psi^{i\alpha} \Psi^j_\alpha,$$
where we combine $\Psi^\alpha, \bPsi_\alpha$ into the quartet  $\Psi^{i\alpha}$. These coupling will preserve the auxiliary nature of superfields
$\Psi$ and will result in a non-flat structure of kinetic terms for spin-variables.
\item Finally, one may add some potential terms for the tensor supermultiplet $V^{ij}$ \cite{IL}.
\end{itemize}
Each of these modifications will result in much more complicated (even in the bosonic sector) systems. It is not clear to us
whether such extensions (or which of them) are interesting. That is why we limited ourselves in the present Letter to the simplest of the possible systems with clear meaning.

Another line of the possible extensions of the present construction is based on the old idea to get a four-dimensional hyper-K\"{a}hler
manifold by dualization of the bosonic auxiliary component in the tensor supermultiplet $V^{ij}$ \cite{{dual},{dual1}}. The first, preliminary
calculations we did, show that in such models there will be strong correlations between the structure of the arising monopoles and the corresponding
hyper-K\"{a}hler metrics. It would be interesting to explore these systems in detail.

\section*{Acknowledgements}
We thank Alexey Isaev, Armen Nersessian and Andrey Shcherbakov for useful discussions. S.K. and A.S. are grateful to the Laboratori Nazionali di Frascati for hospitality.
This work was partially supported by the grants RFBF-08-02-90490-Ukr, 09-02-01209 and 09-02-91349, as well as by the ERC Advanced
Grant no. 226455, \textit{``Supersymmetry, Quantum Gravity and Gauge Fields''%
} (\textit{SUPERFIELDS}).

\def\theequation{A.\arabic{equation}}
\setcounter{equation}0
\section*{Appendix: On-shell realization of N=4 supersymmetry}
The explicit $N=4$ supersymmetry transformations of the physical components leaving the action \p{finaction} invariant read
\bea\label{susycomp}
\delta  v^a &=& - \frac{2i}{x \sqrt{G}}\, \eps^i \bigg [\xi_i v^a + i \eps^{abc} (\sigma^c)^j_i \xi_j v^b \bigg ]
+ \frac{2i}{x \sqrt{G}}\, \bar \eps^i \bigg [ \bar \xi_i v^a + i \eps^{abc} (\sigma^c)^j_i \bar \xi_j v^b \bigg ],\\
\delta \omega^i &=& \frac{i}{x \sqrt{G}} \bigg [\eps^k \omega^i - 2 \eps^i \omega^k \bigg ] \xi_k
+ \frac{i}{x \sqrt{G}} \bigg [\bar \eps_k \omega^i - 2 \bar \eps^i \omega_k \bigg ] \bar \xi^k, \nn\\
\delta \bar \omega_i &=&  \frac{i}{x\sqrt{G}} \bigg [\eps^k \bar \omega_i - 2 \eps_i \bar \omega^k \bigg ] \xi_k
+ \frac{i}{x \sqrt{G}} \bigg [\bar \eps_k \bar \omega_i - 2 \bar \eps_i \bar \omega_k \bigg ] \bar \xi^k, \nn\\
\delta \xi^i &=& -\frac{i G'}{2(G)^{3/2}}\, \eps^i \xi^2 + \frac{x^3}{2} \sqrt{G}\, \bar \eps^j
\bigg [\delta^i_j v^a \dot v^a - i \eps^{abc}(\sigma^c)^i_j v^a \dot v^b \bigg]
+ \frac{2i}{x \sqrt{G}}\, \bar \eps^j \xi_j \bar \xi^i \nn\\
&+& \frac{2i}{x \sqrt{G}}\, \bigg[ 1 + \frac{x}{2} \frac{G'}{G} \bigg ] \bar \eps^j \xi^i \bar \xi_j
+ 2\frac{x}{\sqrt{G}}\, \bigg[ 1 + \frac{x}{2} \frac{G'}{G} \bigg ] \bar \eps^j v^a v^b (\sigma^a)^i_j \Sigma^b
- \frac{x}{ \sqrt{G}} \bar \eps^j v^a v^b (\sigma^a)^i_j I^b,
\nn\\
\delta \bar \xi_i &=&  -\frac{i G'}{2(G)^{3/2}}\, \bar \eps_i \bar \xi^2 + \frac{x^3}{2} \sqrt{G}\, \eps_j
\bigg [\delta^j_i v^a \dot v^a + i \eps^{abc}(\sigma^c)^j_i v^a \dot v^b \bigg]
+ \frac{2i}{x \sqrt{G}}\, \eps_j \bar \xi^j \xi_i \nn\\
&+& \frac{2i}{x \sqrt{G}}\, \bigg[ 1 + \frac{x}{2} \frac{G'}{G} \bigg ] \eps^j \xi_j \bar \xi_i
- 2 \frac{x}{\sqrt{G}}\, \bigg[ 1 + \frac{x}{2} \frac{G'}{G} \bigg ] \eps_j v^a v^b (\sigma^a)^j_i \Sigma^b
+ \frac{x}{ \sqrt{G}} \eps_j v^a v^b (\sigma^a)^j_i I^b. \nn
\eea

\end{document}